\documentstyle[aps,prl,twocolumn,psfig,floats]{revtex}

\begin{document}

\draft

\wideabs{

\title{Tailoring of vibrational state populations with 
light-induced potentials in molecules}

\author{M. Rodriguez$^{1}$\cite{endnote}, K.-A.~Suominen$^{1}$, and 
B. M.~Garraway$^{2}$}

\address{$^1$Helsinki Institute of Physics, PL 9,
FIN-00014 Helsingin yliopisto, Finland\\
$^2$SCOAP \& Centre for Theoretical Physics,
CPES, University of Sussex, Falmer, Brighton, BN1 9QJ, United Kingdom}

\date{\today}

\maketitle

\begin{abstract}
We propose a method for achieving highly efficient transfer between the
vibrational states in a diatomic molecule. The process is mediated by 
{\em strong} laser pulses and can be understood in terms of light-induced 
potentials. In addition to describing a specific molecular system, our results 
show how, in general, one can manipulate the populations of the different 
quantum states in double well systems.
\end{abstract}

}

\narrowtext

Quantum control of molecular processes has recently opened new possibilities for
controlling chemical processes, and also for understanding the multistate
quantum dynamics of molecular systems. With strong and short laser pulses one
can create quantum superpositions of vibrational and continuum states, i.e.,
wave packets, which often propagate like classical objects under the influence
of molecular electronic potentials~\cite{Garraway95}. Typically in
spectroscopy one induces with laser light population transfer between 
individual vibrational states, and describes the process using Franck-Condon 
factors, perturbation theory and a very truncated Hilbert space for the 
molecular wave functions. If the laser light comes in as a very short pulse 
we expect to couple many vibrational states because of the broad spectrum.
However, in this Letter we show that it is possible to start from a single 
vibrational state, and end up selectively on another single vibrational state, 
with very high efficiency, even when one uses strong and fast laser pulses.

We have previously discussed how one can transfer the vibrational
ground state population of one electronic state to the vibrational
ground state of another electronic state~\cite{Garraway98}. This
process was called adiabatic passage in light-induced potentials
(APLIP). These potentials, which depend on time, due to the time
dependence of the laser pulse envelopes, provide a useful description
of the process (see e.g. Ref.~\cite{Giusti95} and references therein). 
Here we use the same kind of light-induced potentials
to describe and understand transfer processes between
excited vibrational states.  However, we will see that the process is more
complicated than the simple adiabatic following assumed in APLIP.

\begin{figure}[tbh]
\centerline{
\psfig{width=85mm,file=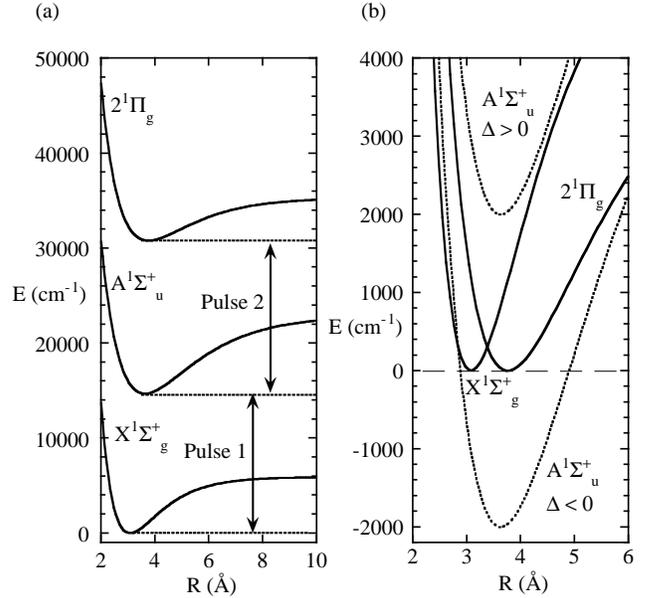}
}
\bigskip
\caption[f1]{(a) The three Na$_2$ potential energy surfaces used in our 
calculations: X$^1\Sigma_g^+$, A$^1\Sigma_u^+$ and 2$^1\Pi_g$.
(b) The shifted potentials. Here we have set
$\Delta=\Delta_1=-\Delta_2$, 
[see Eq.~(\ref{U})] and
show the position of the intermediate A$^1\Sigma_u^+$ state (dotted line)
in respect to the other two electronic states (solid lines) for
$\hbar\Delta=\pm 2000$ cm$^{-1}$. 
\label{pots}}
\end{figure}

To demonstrate the new process we have chosen the same Na dimer potentials, 
shown in Fig.~\ref{pots}(a), as in our previous study~\cite{Garraway98}. 
The three electronic states are coupled in a ladder formation by two laser 
pulses. Instead of describing the system in terms of the vibrational states 
and Franck-Condon factors, we 
consider only the electronic state wave functions, $\Psi_i(R,t)$, $i=X,A,\Pi$. 
If we apply the rotating wave approximation, we can `shift' the $X^1\Sigma_g^+$ 
and $2^1\Pi_g$ state potentials in energy by the corresponding
laser photons. Then we obtain the situation shown in Fig.~\ref{pots}(b),
after a suitable redefinition of the energy zero point. The
resonances between the electronic state potentials now become curve crossings.
The evolution of the three wave functions $\Psi_i(R,t)$ is given by the
time-dependent Schr\"odinger equation with the Hamiltonian
\begin{equation}
   H = -\frac{\hbar^2}{2m}  \frac{ \partial^2 }{ \partial R^2 }\
       {\cal I} + {\cal U}(R,t)
   \label{ham}
\end{equation}
where $R$ is the internuclear separation, $m$ is the reduced mass of the 
molecule, and the electronic potentials and couplings are given by
\begin{equation}
   {\cal U}(R,t) = \left[
   \begin{array}{ccc}
      U_X(R) & \hbar\Omega_1(t)  & 0 \\
      \hbar\Omega_1(t)  & U_A(R)+\hbar\Delta_1 & \hbar\Omega_2(t) \\
      0 & \hbar\Omega_2(t)   & U_\Pi(R)+\hbar(\Delta_1+\Delta_2)
   \end{array} \right].
   \label{U}
\end{equation}
Here $U_X(R)$, $U_A(R)$, and $U_\Pi(R)$ are the three potentials, $\Delta_1$
and $\Delta_2$ are the detunings of the two pulses from the lowest points of
the potentials [dashed lines in Fig.~\ref{pots}(a)], 
and $\Omega_1(t)=\mu_{XA} E_1(t)/\hbar$, $\Omega_2(t)=\mu_{A\Pi}
E_2(t)/\hbar$ are the two Rabi frequencies. We have assumed for simplicity that
the two dipole moments are independent of $R$ and we have used Gaussian pulse
shapes, $\Omega_i(t)=\Omega\exp\{-[(t-t_i)/T]^2\}$, $i=1,2$.

\begin{figure}[tbh]
\centerline{
\psfig{width=85mm,file=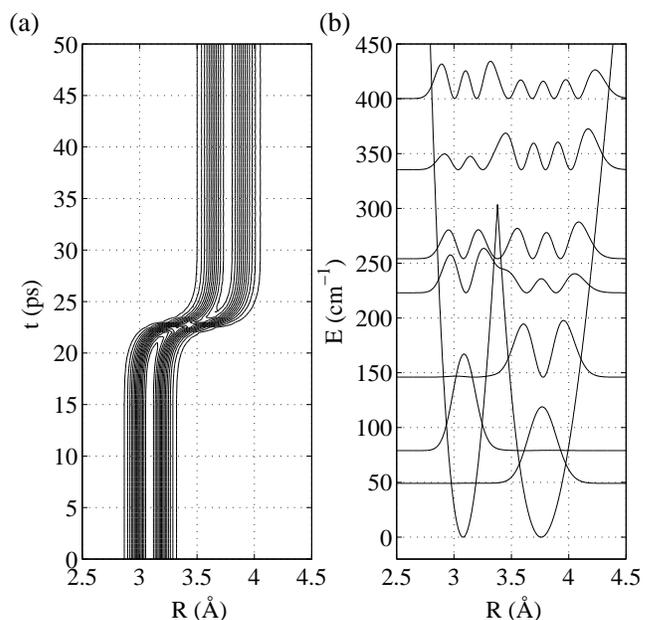}
}
\caption[f2]{(a) The APLIP process for $\hbar\Delta=-3000$ cm$^{-1}$,
$\hbar\Omega=2000$ cm$^{-1}$, $t_1=25.5$ ps, $t_2=20.5$ ps, and $T=5.5$ ps.
(b) The active light-induced state (independent of $\Delta$ at low energies)
when the pulses are absent ($\Omega=0$). 
We also show the seven lowest vibrational
eigenstates $|\phi_i(R)|^2$ of the potential. The left well corresponds to 
X$^1\Sigma_g^+$ and the right one to 2$^1\Pi_g$.
\label{pots2}}
\end{figure}

The light-induced potentials are obtained by diagonalising the potential term
(\ref{U}). For these potentials the curve crossings become avoided crossings.
It is easy to see from Fig.~\ref{pots}(b) that in the absence of the pulses
(or when they both are weak) one of the light-induced states corresponds,
at low energies, to the double well structure formed by the $X^1\Sigma_g^+$ 
and $2^1\Pi_g$ electronic state potentials. In Fig.~\ref{pots}(b) this 
corresponds to the lower part of the two solid curves and we will call this 
the active eigenstate. In Ref.~\cite{Garraway98} we showed that, if the
pulses are applied in a counterintuitive order ($t_1>t_2$), the double well 
structure will disappear as the bottom of the (initially empty) well
on the right moves up and vanishes. (In Fig.~\ref{pots}(b) this is
seen to be because of a repulsion between the $2^1\Pi_g$ and
A$^1\Sigma_u^+$ ($\Delta<0$) state). After this the left 
well broadens and moves to the right. Finally the double well structure
is re-established as the pulses reduce in intensity. 
If we now consider that the ground vibrational state
of the left well is initially populated, that population would follow the
light-induced potential and be transformed smoothly into the ground state wave
function of the right well. This is APLIP, and the smooth change in
the total wave function is
demonstrated in the contour plot in Fig.~\ref{pots2}(a).

If we choose $\Delta_1=-\Delta_2\equiv \Delta$, the bottoms of the two wells are
on the same level initially and finally [as in Fig.~\ref{pots}(b)]. 
The sign of $\Delta$ determines the
evolution of the light-induced potential. For negative $\Delta$ we obtain the
APLIP situation, but for positive $\Delta$ the right well drops down at
first, instead of disappearing. In this case the APLIP situation is
not obtained. 

One is tempted to associate the APLIP process with STIRAP~\cite{Bergmann98}, 
and consider the active light-induced potential as a dark 
state~\cite{Arimondo96}, which, because of the counterintuitive pulse
order, remains 
uncoupled from the other states. It is true that during the APLIP process the 
intermediate state ($A^1\Sigma_u^+$) population
remains very small. However, the process is not the same as STIRAP, 
because the strength of the atom-light coupling, i.e.\ Rabi frequency,
becomes much larger than the vibrational spacing; it is not possible
to isolate only a few energy levels. Furthermore, in a STIRAP process
there would not be a smooth transport of the wave packet in 
Fig.~\ref{pots2}(a) because only two simple vibrational eigenfunctions
would be involved.

STIRAP is also considered to be an adiabatic process, which is not
strictly the case for Fig.~\ref{pots2}.  We can see this by
considering the initial (and final) vibrational states of the
active light-induced potential which matches the two original
electronic potentials as shown in Fig.~\ref{pots2}(b).  The lowest
states correspond to the individual vibrational states of the original
potentials. In STIRAP we should start with the lowest state on left,
and we would expect it to evolve adiabatically into the lowest state
on the right, though not in the smooth way seen for the APLIP process
in Fig.~\ref{pots2}(a).
However, if the evolution were {\em truly} adiabatic the population
would remain in the left well; by definition of adiabatic following it
cannot change state. This is because, for the example shown in
Fig.~\ref{pots2}(b), as the right well rises up, and the
right-hand-well wave function with it, there is a diabatic crossing of
the energy states which allows the population to remain in the left
well as the right well vanishes. As long as the barrier between the
two wells is thick enough, the crossing of vibrational eigenenergies
remains diabatic (uncoupled) and the wave packet (left-hand well
vibrational ground state) can be channelled from left to right
(APLIP). There always has to be at least one such diabatic crossing,
but if there is only one, it could come at the start, or the end, of
the time evolution depending on whether the initial or final
vibrational state is lower in energy. Thus the interesting processes
mediated by the light-induced potentials are different from STIRAP, and they
can not be obtained by perfect adiabatic following. 

\begin{figure}[tbh]
\centerline{
\psfig{width=85mm,file=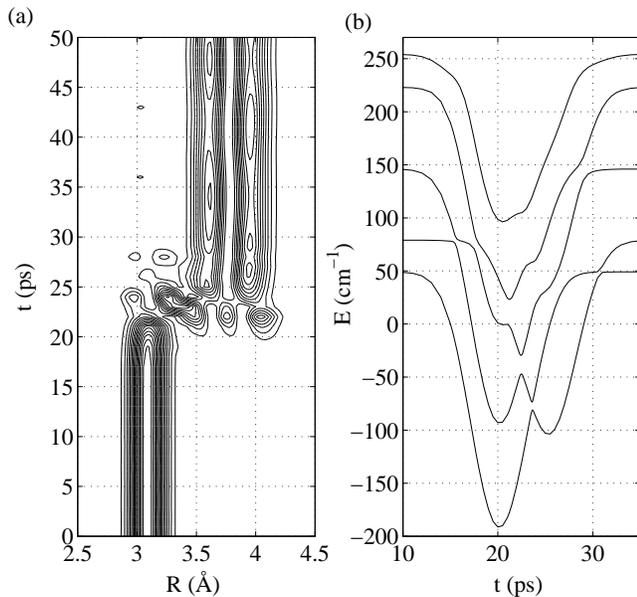}
}
\caption[f3]{
(a) The vibrational state tailoring with $\hbar\Delta=2010$ cm$^{-1}$,
$\hbar\Omega=733$ cm$^{-1}$, $t_1=25.5$ ps, $t_2=20.5$ ps, and $T=5.5$ ps. 
Here we show the time evolution of the total probability distribution
$|\Psi(R,t)|^2=\sum_i|\Psi_i(R,t)|^2$. 
(b) The time-dependent eigenenergies of the five lowest vibrational states of
the active light-induced potential for the situation shown in
(a). [Note that for $t\lesssim 15$ ps and $t\gtrsim 35$ ps 
the eigenstates match those seen in Fig.~\ref{pots2}(b).]
\label{pots3}}
\end{figure}

Figure~\ref{pots2}(b) also displays the initial and final situation if
we now switch to a positive detuning. We choose $\hbar\Delta=2010$~cm$^{-1}$ 
for the time evolution shown in Fig.~\ref{pots3}(a) where $\hbar\Omega= 
733$~cm$^{-1}$ and $t_1-t_2=T=5.5$~ps. We denote the vibrational quantum 
number of the states in the electronic potential with $\nu$, and indicate with 
the terms ``left well'' and ``right well'', whether  these states
correspond to the initial state 
($X^1\Sigma_u^+$), or the final state ($2^1\Pi_g$), respectively. We 
transfer the $\nu=0$ state on the left well into the 
$\nu=1$ state on the right well. The process is very efficient: the occupation 
probability of the right well ($2^1\Pi_g$ state) in the end is 94 \%. Clearly
the crucial moment is around $t=20$~ps where the original single-peak
distribution extends to the right and forms temporarily a four-peaked
distribution over the combined well system. Also, around $t=25$~ps
the four-peak structure compresses into the two-peak structure located on the
right well. 

In order to understand the process, we have calculated the time
dependence of the eigenenergies of the active light-induced potential
[seen for initial and final time in Fig.~\ref{pots2}(b)].  The figure
shows clearly that if the evolution is fully adiabatic, no change of
$\nu$ can occur during the time evolution because there are no crossings of  
eigenenergies. This also means that no change of well is possible either.
However, we do see several avoided crossings between the eigenenergies and 
the point is that when the pulses are weak, these crossings are passed
diabatically, and when the pulses are strong they are passed adiabatically.

\begin{figure}[tbh]
\centerline{
\psfig{width=85mm,file=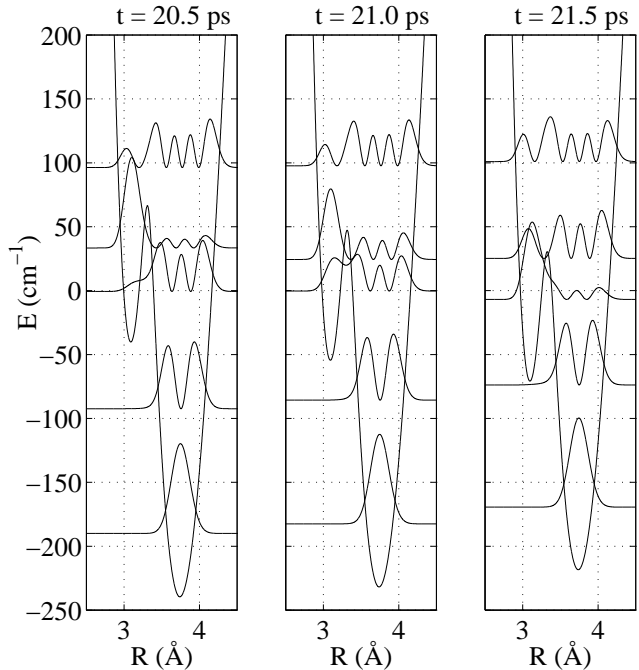}
}
\caption[f4]{
The behaviour of the eigenstates of the active light-induced potential in the
situation of Fig.~\ref{pots3}. We see how the fourth eigenstate deforms within
1 ps from the initial single-peak structure into the four-peak one. An inverse
process happens to the third eigenstate. In the diabatic situation the system
would follow the single-well eigenstates, thus jumping from the fourth state
to the third. However, here the system instead follows the fourth state 
nearly adiabatically.
\label{pots4}}
\end{figure}

In our example, Fig.~\ref{pots3}(b), the second lowest state
corresponds initially, and finally, to the lowest vibrational state of
the left well (i.e., the state populated initially in our calculation)
as may be confirmed by inspecting Fig.~\ref{pots2}(b). During the time evolution
the wave function passes the two first avoided
crossings diabatically, but for $t>20$ ps it follows nearly adiabatically the
fourth eigenstate. Just before $t=30$ ps it moves diabatically onto the third
eigenstate, which asymptotically corresponds with the $\nu=1$ state on the
right well [see again Fig.~\ref{pots2}(b)].  In
Fig.~\ref{pots4} we show the change at the crucial moment around $t=21$ ps. 
The transition from a narrow single well wave function into a wide combined
two-well wave function takes place very quickly. A fascinating aspect is that 
the evolution of the total wave function still follows the fourth vibrational 
eigenstate of the active light-induced potential. A similar adiabatic
following happens also near $t=25$ ps when the third and the fourth eigenstate
again approach each other.

\begin{figure}[tbh]
\centerline{
\psfig{width=85mm,file=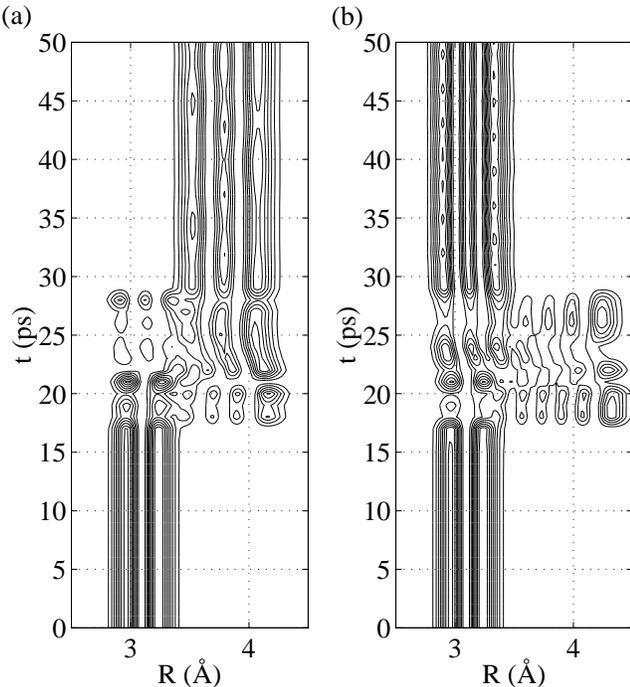}
}
\caption[f5]{
The state tailoring process with $\nu=1$ in the left well as the initial state.
(a) $\nu=2$ in the right well as the final state. Here $\hbar\Delta_1=2010$ 
cm$^{-1}$ and $\hbar\Delta_2=-1960$ cm$^{-1}$.
(b) $\nu=2$ in the left well as the final state. Here $\hbar\Delta_1=2010$ 
cm$^{-1}$ and $\hbar\Delta_2=-2160$ cm$^{-1}$.
\label{pots5}}
\end{figure}

The example given here is not unique. By allowing $\Delta_1\neq -\Delta_2$ we
have a very good handle in choosing which single well states are paired into a
combined well state. In Fig.~\ref{pots5} we show two other examples. They
demonstrate that the process does not require the initial state on the left to
be $\nu=0$, and the process can be used also to change $\nu$ but keep the
initial and final well the same~\cite{extension}. 

Even in the region where both pulses are strong the adiabatic following of a
single vibrational eigenstate is not complete. Small oscillating contributions
appear in the final state. The contour plots tend to emphasise these
oscillations. For example, in the particular case of Fig.~\ref{pots3}(a), 
only about 2 \% probability of diabatic transfer creates these oscillations.

The examples shown here for tailoring the vibrational state populations have
two advantages over the APLIP process. Firstly, having $\Delta>0$ (a region in 
which APLIP does not work) moves the central frequencies of the two laser
pulses away from each other. Secondly, the required intensities are clearly
smaller as we mainly need to shift the potentials, but not to deform them
strongly. Of course, the discussion in Ref.~\cite{Garraway98} regarding the
role of electronic states outside our truncated three-state system, and the
rotational states still apply. We have used Gaussian pulse shapes, but with
other pulse shapes it might be possible to achieve even better control of the 
adiabaticity at different avoided crossings between the eigenstates.

In this Letter we have shown how, even with strong and short laser pulses one
can still perform very selective and yet efficient tailoring of vibrational
state populations in molecules. If 
we were to describe the process using the vibrational state basis
of the three electronic potentials, the treatment would involve
complicated transfer processes between a large number of states. Our
presentation shows that the process can be understood very well in terms of the
light-induced potentials, and their time-dependent vibrational eigenstates.
Furthermore, this description also allows one to identify how the system
parameters should be set in order to achieve specific outcomes. The main
ingredient is establishing the right balance between the adiabatic following of
the time-dependent vibrational eigenstates, and nonadiabatic transfer between
them at avoided crossings. We have discussed the situation in a molecular
multistate framework, but in the light-induced potential description the 
relevant process really takes place within a general two-well potential 
structure. Thus
our observations hold also for any two-well structure, where the barrier
height and the well depths can be controlled time dependently in similar
manner.

This work was supported by the Academy of Finland, Project no.~43336.


\begin{references}

\bibitem[*]{endnote} Present address: Department of Electrical Engineering,
Helsinki University of Technology, PL 9400, FIN-02015 TKK, Finland

\bibitem{Garraway95} B. M. Garraway and K.-A. Suominen, 
                     Rep. Prog. Phys. {\bf 58}, 365 (1995).

\bibitem{Garraway98} B. M. Garraway and K.-A. Suominen,
                     Phys. Rev. Lett. {\bf 80}, 932 (1998).

\bibitem{Giusti95}   A. Giusti-Suzor, F. H. Mies, L. F. DiMauro, E. Charron,
                     and B. Yang, J. Phys. B {\bf 28}, 309 (1995).

\bibitem{Bergmann98} K. Bergmann, H. Theuer, and B. W. Shore,
                     Rev. Mod. Phys. {\bf 70}, 1003 (1998).

\bibitem{Arimondo96} E. Arimondo, 
                     in {\it Progress in Optics XXXV}, 
                     ed.~E. Wolf (North-Holland, Amsterdam, 1996), p. 257.

\bibitem{extension}  A rather obvious extension of APLIP is to start with a
                     $\nu>0$ state, and move from the left well to the right 
                     one while preserving $\nu$. 

\end{references}
\end{document}